\documentclass[10pt]{article}
\usepackage{amsmath}
\usepackage{amssymb}
\usepackage{multicol}
\newcommand{\we}{{\,\wedge\,}}
\newcommand{\tn}{{\,\otimes\,}}
\newcommand{\Tcal}{{\mathcal{T}}}
\newcommand{\Ucal}{{\mathcal{U}}}
\newcommand{\ee}{{\mathsf{e}}}
\newcommand{\xx}{{\mathsf{x}}}
\newcommand{\dx}{\dO\xx}
\newcommand{\de}{\partial}
\newcommand{\dde}[2]{\frac{\partial #1}{\partial #2}}
\newcommand{\nasl}{{\rlap{\raise1pt\hbox{\,/}}\nabla}}
\newcommand{\e}{\varepsilon}
\newcommand{\brth}{\breve\theta}
\newcommand{\Ii}[2]{{}^{#1}_{\phantom{#1}\!#2}}
\newcommand{\iI}[2]{{}_{#1}^{\phantom{#1}\!#2}}
\newcommand{\iIi}[3]{{}_{#1\phantom{#2}\!\!#3}^{\phantom{#1}\!#2}}
\newcommand{\sI}{{\scriptscriptstyle I}}
\newcommand{\dO}{\mathrm{d}}
\newcommand{\iO}{\mathrm{i}}

\newcommand{\spec}[1]{{}_{\scriptscriptstyle{\mathrm{#1}}}}
\newcommand{\Tr}{\operatorname{Tr}}

\title{~\\[-48pt]{\Large On the notions of energy tensors
in tetrad-affine gravity}}
\author{\\[-18pt]Daniel Canarutto \\
{\small\it Dipartimento di Matematica e Informatica ``U.~Dini'', }\\[-3pt]
{\small\it Via S. Marta 3, 50139 Firenze, Italia}\\[-3pt]
{\small email:~daniel.canarutto@unifi.it}\\[-3pt]
{\small http://www.dma.unifi.it/\char126 canarutto}}
\date{{~\\[-8pt]\small November 16, 2017 (accepted version)} }
\begin{document}

\maketitle
\begin{abstract}
We are concerned with the precise modalities by which
mathematical constructions related to energy-tensors
can be adapted to a tetrad-affine setting.
We show that, for fairly general gauge field theories
formulated in that setting,
two notions of energy tensor|the canonical tensor
and the stress-energy tensor|exactly coincide with no need for tweaking.
Moreover we show how both notions of energy-tensor can be naturally extended
to include the gravitational field itself,
represented by a couple constituted by the tetrad and a spinor connection.
Then we examine the on-shell divergences of these tensors
in relation to the issue of local energy-conservation in the presence of torsion.
\end{abstract}
\bigbreak

\bigbreak
PACS 2010: 02.40.-k, 
04.20.Cv, 
04.20.Fy. 
\smallbreak

\begin{multicols}{2}
\subsubsection*{Introduction}

In Lagrangian field theory~%
\cite{GS73,GotayMarsden92,Kuperschmidt79,LandauLifchitz68,MaMo83b,Tr67}
one has a precise mathematical construction yielding a `canonical energy-tensor'
associated with each field sector.
Such tensors are related to conservation laws by generalizations
of the classical Noether theorem,
which constitutes the basis for physical interpretation.
When the considered field theory is formulated over a curved Lorentzian
background then one has the further notion of
`stress-energy tensor'~\cite{LandauLifchitz68,FerrarisFrancaviglia92,
FatibeneFerrarisFrancaviglia94,ForgerRoemer04},
whose relation with the canonical energy-tensor is known as the
`Belinfante-Rosenfeld formula'~\cite{Belinfante40,Rosenfel40,GotayMarsden92}.
In various concretely interesting cases the two said notions yield
tensors which turn out to be different just by a numerical coefficient and,
possibly, by a needed symmetrization.

In particular,
the notion of energy tensor for the gravitational field
has been variously debated in the literature~%
\cite{BambaShimizu15,ChenNester00,KijowskiGRG97,Komar59,Lauloiso08,Leclerc06}.
Recent results~\cite{Padmanabhan1312.3253} suggest that that role
should be played by the Ricci tensor.
On the other hand, precise covariant constructions~\cite{C16c} show
that the Ricci tensor is to be seen
as the canonical tensor of the gravitational field.

In this paper we are interested in applying the general formalism
of Lagrangian field theory in the context where
a gauge theory is coupled with tetrad-affine gravity|indeed we regard that
as the most natural and convenient setting.
This also yields a canonical tensor for the gravitational field that, again,
turns out to be essentially the Ricci tensor.
It should be stressed, however, that
we do not aim at a detailed discussion of the possible physical interpretations
of the ensuing mathematical notions,\footnote{
Indeed, a straightforward physical interpretation of the Ricci tensor
in terms of energy is problematic as, for example, Schwarzschild spacetime has
non-zero gravitational energy while the Ricci tensor vanishes.}
which are introduced just as natural extensions of usual notions.

Tetrad gravity~\cite{Cartan1922,Cartan1923_25,Hehl1973_I,Hehl1974_II,
HehlHeydeKerlickNester76,HCMN95,Henneaux78,Kibble61,Poplawski09,Sciama54,Trautman06}
has been introduced and studied mainly as a convenient
`non-holonomic coordinate' formalism,
but it is interesting to note that the tetrad $\theta$
acquires a neat geometric meaning
if it is viewed as an isomorphism between the tangent bundle $\mathrm{T}M$
of the spacetime manifold $M$ and a further vector bundle $H$ over $M$
whose fibers are endowed with a Lorentz metric $g$|i.e.\ an
$\mathrm{SO}(1,3)$-bundle.
Moreover such $H$ is naturally generated by the spinor bundle
needed for the description of Dirac fields,
so that it does not actually constitute an \emph{ad hoc} unphysical assumption;
this result is specially well expressed
in the context of 2-spinor geometry~\cite{PR84,PR88,C98,C00b,C07,C10a,C15b}.
Now $\theta$ transforms $g$ into a spacetime metric;
moreover a metric connection $\Gamma$ of $H$ is transformed by $\theta$
into a metric spacetime connection.
Thus the couple $(\theta,\Gamma)$ can be regarded
as representing the gravitational field,
according to what we may call the `tetrad-affine representation'.
Note that the spacetime structures, in this view,
are derived, non-fundamental quantitites.
Though the spacetime metric also determines
the Levi-Civita (symmetric) connection,
the spacetime connection corresponding to $\Gamma$ has non-zero torsion,
which turns out to interact with spin fields.
Torsion is then unavoidable, but not a fundamental field,
since it can be essentially expressed as the covariant differential\footnote{%
The notion of covariant differential of vector-valued forms,
which has been variously present in the literature for several years,
is strictly related to the
\emph{Fr\"olicher-Nijenhuis bracket}~%
\cite{FroNij56,MaMo83b,ManMod84,KolarMichorSlovak93,C16e}.
In this paper we will just write down the needed coordinate expressions.} %
of $\theta$ with respect to $\Gamma$.
Furthermore, we observe that $\Gamma$ can be essentially regarded as the
spinor connection, as shown by eq.~\eqref{eq:GammafromCs}.

In general, field theory topics
can be most rigorously addressed in the context of a formulation exploiting
jet bundle geometry~\cite{Gar74,GS73,Krupka15,Kuperschmidt79,MaMo83b,Sa89,Tr67}.
In this presentation, however, we will skip some technicalities of that kind,
limiting ourselves to plain coordinate expressions,
even though a mathematically exigent reader might regard some statements
as not sufficiently justified.

\subsubsection{Tetrad-affine gravity}

If $\bigl(\ee_\lambda\bigr)$ is an orthonormal frame of $H$
then the tetrad can be expressed as
\hbox{$\theta=\theta_a^\lambda\,\dx^a\tn\ee_\lambda$}\,,
where the components $\theta_a^\lambda$ have the physical dimension of a length.
We will use shorthands
\begin{align*}
&|\theta|\equiv\det\theta=\tfrac1{4!}\,
\e^{abcd}\,\e_{\lambda\mu\nu\rho}\,
\theta_a^\lambda\theta_b^\mu\theta_c^\nu\theta_d^\rho~,
\displaybreak[2]\\[6pt]
&\brth^a_\lambda\equiv\de|\theta|/\de\theta_a^\lambda= \tfrac1{3!}\,
\e^{abcd}\,\e_{\lambda\mu\nu\rho}\,\theta_b^\mu\theta_c^\nu\theta_d^\rho~,
\displaybreak[2]\\[6pt]
&\brth^{ab}_{\lambda\mu}\equiv\de\brth^a_\lambda/\de\theta_b^\mu=
\tfrac12\,\e^{abcd}\,\e_{\lambda\mu\nu\rho}\,\theta_c^\nu\theta_d^\rho~,
\displaybreak[2]\\[6pt]
&\brth^{abc}_{\lambda\mu\nu}\equiv\de\brth^{ab}_{\lambda\mu}/\de\theta_c^\nu=
\e^{abcd}\,\e_{\lambda\mu\nu\rho}\,\theta_d^\rho~.
\end{align*}
We observe that the above quantities
are well-defined also if $\theta$ is degenerate;
if $\theta$ is invertible then
\hbox{$(\theta^{-1})^a_\lambda=\brth^a_\lambda/|\theta|$}\,.

We denote the components of the metric and of a connection of $H$
by $g_{\lambda\mu}$ and $\Gamma\!\iIi a\lambda\mu$\,, respectively,
and the induced spacetime quantities by
\begin{equation}
g_{ab}\equiv\theta_a^\lambda\theta_b^\mu\,g_{\lambda\mu}~,\qquad
\Gamma\!\iIi acb=
\theta^c_\lambda\,(-\de_a\theta_b^\lambda+\Gamma\!\iIi a\lambda\mu\,\theta_b^\mu)~,
\end{equation}
where \hbox{$(\theta^{-1})^a_\lambda=
\theta^a_\lambda\equiv g^{ab}\,g_{\lambda\mu}\,\theta_b^\mu$}\,.
Then $\theta$ can be regarded as a `square root of the metric',
and we also get \hbox{$|\theta|\equiv\det\theta=\sqrt{|\det g|}$}\,.
The condition that the tetrad be covariantly constant characterizes
a connection of the spacetime manifold which turns out to be metric,
but does not coincide with the standard spacetime connection
since it is not symmetric 
(remark: for the connection coefficients we use the sign convention
yielding \hbox{$\nabla\!_a\dx^c=\Gamma\!\iIi acb\,\dx^b$}).
The torsion is expressed as
\begin{equation}
T\Ii c{ab}=\Gamma\!\iIi bca-\Gamma\!\iIi acb=
\theta^c_\lambda\,(\de_{[a}^{\phantom{\mu}}\theta_{b]}^\lambda
+\theta_{[a}^\mu\,\Gamma\!\iIi{b]}\lambda\mu)~.
\end{equation}

Locally, we write the Lagrangian density of a field theory
as $\ell\,\dO^4\xx$\,,
where $\ell$ is a function of the fields and their first derivatives.
For the gravitational field we set
\begin{equation}
\ell\spec{grav}=\tfrac{1}{4G}\,R\iI{ab}{\lambda\mu}\brth^{ab}_{\lambda\mu}=
-\tfrac{1}{2G}\,R\iI{ab}{\lambda\mu}\theta^b_\lambda\theta^a_\mu\,|\theta|~,
\end{equation}
where\footnote{%
Here ${\scriptstyle G}$ is Newton's gravitational constant.
We use natural units: \hbox{$\hbar=c=1$}\,.}
\begin{equation*}
R\iI{ab}{\lambda\mu}\equiv R\iIi{ab}\lambda\nu\,g^{\nu\mu}=
(-\de^{\phantom{A}}_{[a}\Gamma\!\iIi{b]}\lambda\nu
+\Gamma\!\iIi{[a}\lambda\rho\,\Gamma\!\iIi{b]}\rho\nu)\,g^{\nu\mu}~.
\end{equation*}
If $\theta$ is non-degenerate then
\hbox{$R\iI{ab}{\lambda\mu}\theta^b_\lambda\theta^a_\mu$} coincides
with the scalar curvature of the spacetime connection,
but note that the above Lagrangian density
is well-defined also in the degenerate case.

Independent variations of the fields
$\theta_a^\lambda$ and $\Gamma\!\iI a{\lambda\mu}$ then yield
the Euler-Lagrange operator components
\begin{align} \label{eq:Einsteintensor}
&(\delta\ell\spec{grav})_\lambda^a=
\tfrac1{4G}\,\brth_{\lambda\mu\nu}^{abc}\,R\iI{bc}{\mu\nu}=
\tfrac1G\,\brth^b_\lambda\,E\iI ba~,
\\[6pt]
&(\delta\ell\spec{grav})\Ii{a}{\lambda\mu}=
-\tfrac1{4G}\,T\Ii e{bc}\,\theta_e^\nu\,\brth^{abc}_{\lambda\mu\nu}~,
\end{align}
where $E\iI ba$ is the Einstein tensor (not symmetric in this context).

\subsubsection{Gauge field theories in tetrad-affine gravity}

A spin-zero `matter field' in a gauge theory
is a section of some vector bundle
whose fibers are not `soldered' to spacetime.
A field with non-zero spin can be seen as a section of a similar bundle
tensorialized by a spin bundle;
we denote its components by $\phi^{i\alpha}$,
where $\alpha$ is the spin-related index
(which may represent a sequence of ordinary spinor indices).
The adjoint field $\bar\phi_{i\alpha}$ can be regarded
as an independent section of the dual bundle.

The matter fields interact with a gauge field $A\iIi aij$
that is a connection of the `unsoldered' bundle.
Ususally $A$ is assumed to preserve some fiber structure
and is accordingly valued into the appropriate Lie algebra,
so one uses components \hbox{$A_a^\sI$}\,,
but we won't need to deal with such restriction explicitely|it is not difficult
to see that the arguments presented here work seamlessly
with respect to the needed restriction.
The covariant derivative of the matter field has the expression
\begin{equation*}
\nabla\!_a\phi^{i\alpha}=\de_a\phi^{i\alpha}
-A\iIi aij\,\phi^{j\alpha}-\omega\!\iIi a\alpha\beta\,\phi^{i\beta}~,
\end{equation*}
where the `spinor connection' $\omega\iIi a\alpha\beta$
is related to $\Gamma\!\iIi a\lambda\mu$
by a linear relation of the type
\begin{equation*}
\omega\iIi a\alpha\beta=G^{\alpha|\mu}_{\beta|\lambda}\,\Gamma\!\iIi a\lambda\mu~.
\end{equation*}
The coefficients $G^{\alpha|\mu}_{\beta|\lambda}$ can be expressed as combinations
of Kronecker deltas in the case of integer spin,
while Dirac matrices are involved for semi-integer spin.
In particular, for spin one-half we have
\begin{equation}\label{eq:spingravGamma}
\omega\iIi a\alpha\beta=\tfrac14\,\Gamma\!\iIi a\lambda\mu\,(\gamma_\lambda\gamma^\mu)\Ii\alpha\beta~,
\end{equation}
which can be inverted as
\begin{equation}\label{eq:GammafromCs}
\Gamma\!\iIi a\lambda\mu=\tfrac12\Tr(\gamma^\lambda\,\omega\!_a\,\gamma_\mu)~.
\end{equation}
Thus our variable $\Gamma$ could be regarded as the spinor connection,
namely the gravitational field can be equivalently represented
as the couple $(\theta,\omega)$.

\smallbreak\noindent{\sc remark}: In this concise exposition,
charges and other factors that usually appear in the literature
are absorbed into the gauge field itself.\smallbreak

The Klein-Gordon Lagrangian, written in the form
\begin{equation}\label{eq:KleinGordonLagrangian}
\ell_\phi=\tfrac1{2|\theta|}\,g^{\lambda\mu}\,\theta^a_\lambda\theta^b_\mu\,
\nabla\!_a\bar\phi_{i\alpha}\,\nabla\!_b\phi^{i\alpha}
-\tfrac12\,m^2\,\bar\phi_{i\alpha}\,\phi^{i\alpha}\,|\theta|~,
\end{equation}
yields the well-defined density $\ell_\phi\,\dO^4\xx$
for any matter field.
For a field of spin one-half one rather uses
\begin{equation}
\ell_\psi=\bigl(
\tfrac\iO2\,(\bar\psi_{\alpha i}\,\nasl\psi^{\alpha i}
-\nasl\bar\psi_{\alpha i}\,\psi^{\alpha i})
-m\,\bar\psi_{\alpha i}\,\psi^{\alpha i}\bigr)\,|\theta|~.
\end{equation}
Note that the Dirac operator \hbox{$\nasl\equiv\gamma^a\nabla\!_a$}
depends on the tetrad,
that transforms the natural Clifford algebra structure of $H$|and its
representation on the Dirac spinor bundle|into an object
defined on $\mathrm{T}M$ .

For matter fields of either integer or semi-integer spin greater than one-half
one may wish to consider an appropriate specialized setting,
leading to possible generalizations
of the Dirac equation~\cite{BargmannWigner48,BarthChristensen83,C16f}.
However, issues about the Lagrangian treatment of such setting
suggest that we provisionally confine ourselves to the
Lagrangian~\eqref{eq:KleinGordonLagrangian}
for all matter fields of spin different from one-half.

A convenient handling of gauge fields,
analogous to the metric-affine gravity formalism,
treats the gauge field $A$ and the tensor field $F$
as independent fields~\cite{C98}.
Indeed, consider the Lagrangian
\begin{equation}
\ell\spec{gauge}=
-\tfrac12\,\brth^{ab}_{\lambda\mu}\,(\dO[A]A)\iIi{ab}ij\,F\Ii{\lambda\mu\,j}i
+\tfrac14\,F\Ii{\lambda\mu\,i}j\,F\!\iIi{\lambda\mu}ji\,|\theta|~,
\end{equation}
where
\hbox{$(\dO[A]A)\iIi{ab}ij=\de_{[a}A\iIi{b]}ij-A\iIi{[a}ih\,A\iIi{b]}hj$}
is the `covariant exterior differential'~\cite{ManMod84,KolarMichorSlovak93}
of $A$\,, coinciding with minus its curvature tensor.
Since $F$ is not present in other pieces of the total Lagrangian
\hbox{$\ell\spec{tot}\equiv\ell\spec{grav}+\ell\spec{matter}+\ell\spec{gauge}$}\,,
with $\ell\spec{matter}$ being either $\ell_\phi$ or $\ell_\psi$\,,
the variation of $\ell\spec{gauge}$ with respect to $F$ immediately yields
\begin{equation}\label{eq:FisdA}
F\!\iIi{ab}ij\equiv\theta_a^\lambda\theta_b^\mu\,F\!\iIi{\lambda\mu}ij=
2\,(\dO[A]A)\iIi{ab}ij~.
\end{equation}

\subsubsection{Energy tensors}

In standard Einstein gravity, the general link between a field's Lagrangian
and the related stress-energy tensor
has non-trivial aspects~\cite{LandauLifchitz68,HE},
mainly since one has to allow for the Lagrangian to depend on the
derivatives of the metric.
In the usual Lagrangians of matter fields this dependance comes from
the spacetime connection coefficients in covariant derivatives,
while the situation is somewhat different in a metric-affine approach.
In the tetrad-affine approach,
the total Lagrangians for all basic cases do not
depend on the derivatives of the tetrad
(later we'll also consider a possible such dependence).
Hence the role of the stress-energy tensor for each sector is played by
\hbox{$\Tcal{}^a_\lambda\equiv\de\ell/\de\theta_a^\lambda=
(\delta\ell)^a_\lambda$}\,.
We obtain
\begin{align} \label{eq:Tgrav}
&(\Tcal\!\!\spec{grav})_\lambda^a=
\tfrac1{4G}\,\brth_{\lambda\mu\nu}^{abc}\,R\iI{bc}{\mu\nu}~,
\\[6pt]
&(\Tcal\!\!\spec{gauge})_\lambda^a=
F\!\iIi{\lambda\nu}ij\,F\Ii{\lambda\mu\,j}i\,\brth^c_\mu
-\tfrac14\,F\!\iIi{\lambda\mu}ij\,F\Ii{\lambda\mu\,j}i\,\brth^c_\nu~,
\\[6pt]
\begin{split}&(\Tcal\!\!_\phi)^c_\nu=
\tfrac1{2|\theta|^2}g^{\lambda\mu}(
\brth^a_\lambda\brth^b_\mu\brth^c_\nu
-\brth^a_\lambda\brth^b_\nu\brth^c_\mu
-\brth^a_\nu\brth^b_\mu\brth^c_\lambda)
\nabla\!_a\bar\phi_{\alpha i}\nabla\!_b\phi^{\alpha i}
\\&\qquad\qquad\qquad\qquad\qquad\qquad\qquad
-\tfrac12\,m^2\bar\phi_{\alpha i}\,\phi^{\alpha i}\brth^c_\nu~,\end{split}
\\[6pt]
\begin{split}&(\Tcal\!\!_\psi)^c_\nu=
\ell_\psi\,\theta^c_\nu-{}
\\&\quad
-\tfrac\iO{2|\theta|}\,g^{\lambda\mu}\,\brth^a_\nu\brth^c_\lambda\,(
\bar\psi_{\alpha i}\,\gamma\iIi\mu\alpha\beta\nabla\!_a\psi^{\beta i}
-\nabla\!_a\bar\psi_{\beta i}\,\gamma\iIi\mu\beta\alpha\psi^{\alpha i})~.\end{split}
\end{align}

Moreover we consider the canonical energy-tensor,
that for a generic field $\phi^i$ has the expression
\begin{equation}
\Ucal\Ii ab=\ell\,\delta\Ii ab-\nabla\!_b\phi^i P\Ii ai~,\qquad
P\Ii ai\equiv \de\ell/\phi^i_{,a}~.
\end{equation}
Note the covariant derivative $\nabla\!_b\phi^i$ above,
in contrast with the ordinary partial derivative $\phi^i_{,b}$
appearing most commonly in the literature.
This modification,
which is necessary for $\Ucal$ to be geometrically well-defined in general,
was introduced by Hermann~\cite{Hermann75};
see also Hehl et al.~\cite{HehlHeydeKerlickNester76},~eq.\;3.10.
A precise geometric construction and a discussion of the meaning of this object
can be found in previous work~\cite{CM85,C16c}.

Briefly, $\Ucal$ relates infinitesimal transformations of the spacetime manifold $M$,
represented by vector fields $X$ on $M$,
to currents of the field theory under consideration,
that are expressed as \hbox{$J^a=\Ucal\Ii ab\,X^b$}\,.
In order to do that one needs a way to `lift' a vector field so that
it acts on the theory's `configuration bundle';
if the latter is not trivial then the required construction can be performed
by means of a connection.
In terms of the coordinate expression of $\Ucal$ this eventually amounts
to replacing $\phi^i_{,b}$ with $\nabla\!_b\phi^i$ in the basic expression.

It is well known that the two notions of energy-tensor
turn out to be strictly related,
though in general they do not coincide~\cite{GotayMarsden92}.
In our present context we can try a generic comparison between $\Tcal$ and $\Ucal$
by observing that writing \hbox{$\ell=\tilde\ell\,|\theta|$}\,,
and assuming that $\ell$ is independent of the derivatives of $\theta$\,,
we get
\begin{align*}
&\Tcal{}^a_\lambda=
\ell\,\theta^a_\lambda+\dde{\tilde\ell}{\theta_a^\lambda}|\theta|~,
\\[6pt]
&\Ucal{}^a_\lambda=\theta^b_\lambda\,\Ucal\Ii ab=
\ell\,\theta^a_\lambda-\theta^b_\lambda\,\nabla\!_b\phi^i P\Ii ai~.
\end{align*}
Then the two tensors coincide if
\begin{equation*}
\dde{\tilde\ell}{\theta_a^\lambda}=
-\frac1{|\theta|}\,\nabla\!_b\phi^i\,P\Ii ai\,\theta^b_\lambda~.
\end{equation*}
Straightforward computations then show that this situation actually occurs
in the basic cases presently under consideration,
including the Dirac spinor case.
Interestingly, this also holds true for the energy tensors
of the gauge and gravitational fields,
provided that we use the right notion of `covariant derivative'
of such fields.
Various arguments~\cite{C16c,C16e} clearly indicate that the role of
the covariant derivative of a connection is to be taken up
by the exterior covariant differential of the connection with respect to itself,
that is minus its curvature tensor. Namely we insert
\hbox{$\nabla\!_bA\iIi cij\equiv(\dO[A]A)\iIi{bc}ij$} into
\begin{equation*}
(\Ucal\spec{gauge})^a_\lambda=
\ell\spec{gauge}\,\theta^a_\lambda
-\theta^b_\lambda\,\nabla\!_bA\iIi cij\,\dde{\ell\spec{gauge}}{(\de_aA\iIi cij)}
\end{equation*}
and obtain the stated identity.
As for the gravitational field $(\theta,\Gamma)$,
since $\ell\spec{grav}$ is independent of the derivatives of $\theta$ we get
\begin{align*}
(\Ucal\!\spec{grav})_\lambda^a&=
\tilde\ell\spec{grav}\,\brth^a_\lambda
-\nabla\!_b\Gamma\!\iI c{\mu\nu}
\dde{\tilde\ell\spec{grav}}{(\de_a\Gamma\!\iI c{\mu\nu})}
\,\brth^b_\lambda\,|\theta|^{-1}=
\\
&= -\tfrac{1}{2G}\,R\,\brth^a_\lambda
-\tfrac{1}{2G|\theta|}\,\brth^{ac}_{\mu\nu}\,R\iI{bc}{\mu\nu}\,\brth^b_\lambda=
\\
&=\tfrac{1}{G}\,(R\iI b\mu\,\theta^a_\mu\,\brth^b_\lambda
-\tfrac12\,R\,\brth^a_\lambda)=(\Tcal\!\!\spec{grav})_\lambda^a~.
\end{align*}

More generally, one may wish to consider a Lagrangian that also depends
on the derivatives of the tetrad.
Then the question arises if one can generalize the construction
of the canonical energy-tensor to this case.
Without being involved in technical details, we state that
two constructions turns out to be legitimate,
the difference between them being the way in which the action of a vector field
on $M$ is properly lifted.
Essentially, both ways eventually lead to an expression of the type
\hbox{$\Ucal\Ii ab=\ell\Ii ab-\mathrm{D}_b\theta_c^\lambda\,P\Ii{a,c}\lambda$}\,,
where $\mathrm{D}_b$ is a suitable differential operator.
One construction yields just
\hbox{$\mathrm{D}_b\theta_c^\lambda=\nabla\!_b\theta_c^\lambda=0$}\,.
More interestingly, the other construction determines $\mathrm{D}_b\theta_c^\lambda$
to be|somewhat similarly to the connection|the covariant differential
of $\theta$\,, that is essentially the torsion. Namely one gets
\begin{equation*}
\Ucal\Ii ab=
\ell\,\delta\Ii ac-P\Ii{a,c}\lambda\,\theta_e^\lambda\,T\Ii e{cb}~,\qquad
P\Ii{a,c}\lambda\equiv\de\ell/\theta^\lambda_{c,a}~.
\end{equation*}

For example one may consider the standard `ghost Lagrangian',
that in terms of the tetrad can be written as
\begin{align*}
&\ell\spec{ghost}\equiv
g^{\lambda\mu}\,\theta^a_\lambda\theta^b_\mu\,
\bar\chi_{\sI,a}\nabla\!_b\chi^\sI\,|\theta|
-\tfrac1{2\xi}\,f_\sI\,f^\sI\,|\theta|~,
\\[6pt]
&f^\sI\equiv
|\theta|^{-1} g^{\lambda\mu}\,
\de_a(\theta^a_\lambda\theta^b_\mu\,|\theta|\,A_b^\sI)~.
\end{align*}
Here $\chi^\sI$ and $\bar\chi_\sI$ are the ghost and anti-ghost fields,
$\xi$ is a constant, and the index ${\scriptstyle I}$
denotes components in the appropriate Lie algebra.
Then the `gauge fixing Lagrangian'
\hbox{$\ell\spec{fix}\equiv-f_\sI\,f^\sI\,|\theta|/(2\xi)$} introduces into
the total canonical energy tensor, constructed in the above described way,
a term which is linear in the torsion.
Similarly the stress-energy tensor gets a term that can be expressed
as the `variational derivative' of $\ell\spec{fix}$ with respect to $\theta$;
on turn this can be expressed in terms of the torsion,
through somewhat intricate computations.

\subsubsection{Field equations}

Besides eq.~\eqref{eq:FisdA}, the field equations obtained from the variations
of $\ell\spec{tot}$ with respect to
$\theta$, $\Gamma$, $A$, $\phi$ and $\bar\phi$ yield,
rexpectively, the gravitational equation, the torsion equation,
the non-Abelian generalization of second Maxwell equation,
and the generalization of either the Klein-Gordon equation or the Dirac equation.

The gravitational equation is
\begin{equation}
0=\Tcal\!\!\spec{tot}\equiv
\Tcal\!\!\spec{grav}+\Tcal\!\!\spec{gauge}+\Tcal\!\!\spec{matter}~,
\end{equation}
where $\Tcal\!\!\spec{matter}$ is either $\Tcal_\phi$ or $\Tcal_\psi$\,.

The other field equations|in a somewhat concise form|can be written
in the K-G case as
\begin{align} \label{feq:phiGamma}
\begin{split}
&0=-\tfrac1{G}\,\brth^{abc}_{\lambda\mu\nu}\,T\Ii e{bc}\,\brth_e^\nu \\
&\qquad\qquad +2g^{ab}\,G^\alpha_\beta{}_{|\lambda\mu}\,
(\bar\phi_{\alpha i}\,\nabla\!_b\phi^{\beta i}
-\nabla\!_b\bar\phi_{\alpha i}\,\phi^{\beta i})~,\end{split}
\displaybreak[2]\\[6pt] \label{feq:phiA}
&0=(\dO[A]{*}F)\Ii{a\,j}i
+\tfrac12\,|\theta|\,g^{ab}\,(\bar\phi_{\alpha i}\,\nabla\!_b\phi^{\alpha j}
-\nabla\!_b\bar\phi_{\alpha i}\,\phi^{\alpha j})~,
\displaybreak[2]\\[6pt] \label{feq:phi}
&0=(\dO[\Gamma\tn A]{*}\!\nabla\bar\phi)_{\alpha i}
+m^2\,\bar\phi_{\alpha i}\,|\theta|~,
\displaybreak[2]\\[6pt] \label{feq:barphi}
&0=(\dO[\Gamma\tn A]{*}\!\nabla\phi)^{\alpha i}+m^2\,\phi^{\alpha i}\,|\theta|~.
\end{align}
Here the ${*}$ stands for the `Hodge isomorphism'
of exterior forms,\footnote{%
Exterior form components $\xi^a$, $\xi^{ab}$ with higher indices
are to be intended relatively to frames $i(\de\xx_a)\dO^4\xx$,
$i(\de\xx_a\we\de\xx_b)\dO^4\xx$ etc.}
namely
\begin{equation*}
{*}F\Ii{ab\,j}i=g^{ac}g^{bd}|\theta|\,F\!\iIi{cd}ij\,,\quad
{*}\!\nabla\phi^{a\,\alpha i}=g^{ab}|\theta|\,\nabla\!_b\phi^{\alpha i}\,,
\end{equation*}
and $\dO[A]$ and $\dO[\Gamma\tn A]$ are the exterior covariant differentials
with respect to the connections indicated between brackets.
A generalized version of the so-called `replacement principle'
states that these differ from the usual `covariant divergences'
by torsion terms~\cite{C16c}.
In fact we have the identities
\begin{align*}
&\nabla\!_a\xi^{a\,i}=(\dO[K]\xi)^i-T\Ii b{ab}\,\xi^{a\,i}~,
\\[6pt]
&2\,\nabla\!_a\xi^{ba\,i}=(\dO[K]\xi)^{b\,i}
-\tfrac12\,\xi^{ac\,i}\,T\Ii b{ac}-\xi^{ba\,i}\,T\Ii c{ac}~,
\end{align*}
where $\xi$ is a \hbox{$(4\,{-}\,r)$}-form (\hbox{$r=1,2$})
valued in a vector bundle and $K$ is a connection of that same bundle.

Eq.~\eqref{feq:phiGamma} is the torsion equation;
eq.~\eqref{feq:phiA} is the `second Maxwell equation';
eqs.~\eqref{feq:phi} and \eqref{feq:barphi} are the `Klein-Gordon equations'
for $\bar\phi$ and $\phi$\,.

In the Dirac case we find the field equations
\begin{align} \label{feq:psiGamma}
&0=-\tfrac1G\,\brth^{abc}_{\lambda\mu\nu}\,T\Ii e{bc}\,\theta_e^\nu
+\tfrac\iO4\,\brth^a_\nu\,\bar\psi_{\alpha i}\,(
\gamma_\lambda\we\gamma_\mu\we\gamma^\nu)\Ii\alpha\beta\,\psi^{\beta i}~,
\\[6pt] \label{feq:psiA}
&0=(\dO[A]{*}F)\Ii{a\,j}i
-\iO\,\brth^a_\lambda\,\bar\psi_{\alpha i}\,
\gamma\Ii{\lambda\alpha}\beta\,\psi^{\beta j}~,
\\[6pt] \label{feq:psi}
&0= -(\iO\,\nasl\bar\psi_{\beta j}+m\,\bar\psi_{\beta j}
+\tfrac\iO2\,\bar\psi_{\alpha j}\tau_\lambda
\gamma\Ii{\lambda\alpha}\beta)\,|\theta|~,
\\[6pt] \label{feq:barpsi}
&0= (\iO\,\nasl\psi^{\beta j}-m\,\psi^{\beta j}\,
+\tfrac\iO2\,\tau_\lambda
\gamma\Ii{\lambda\beta}\alpha \psi^{\alpha j})\,|\theta|~,
\end{align}
where \hbox{$\tau_\lambda\equiv\theta_\lambda^a\,T\Ii b{ab}$}\,.
Eqs.~\eqref{feq:psi} and \eqref{feq:barpsi} are the Dirac equations with torsion.

\subsubsection{Divergences}

In the standard, torsion-free formulation of General Relativity,
the stress-energy tensor in
the right-hand side of the Einstein equation is divergence-free on-shell
(namely when the field equations are taken into account).
This well-known result~\cite{LandauLifchitz68,HE} is a consequence
of the naturality of the Lagrangian,
and holds in particular for gauge theories
provided that the stress-energy tensor contains the contributions
of the matter field \emph{and} the gauge field~\cite{C16c}.
This property is interpreted as local energy-conservation.\footnote{
As previously observed, in this paper we are not involved with
detailed discussions about physical interpretations of the presented
mathematical notions.}

In the presence of torsion the situation is more intricate.
The gravitational equation \hbox{$\Tcal\!\!\spec{tot}=0$} implies
\hbox{$\nabla\!_a(\Tcal\!\!\spec{tot})^a_\lambda=0$}\,,
but the single contributions have non-vanishing divergence.
In particular we remark that the `Einstein tensor' appearing
in eq.~\eqref{eq:Einsteintensor} and eq.~\eqref{eq:Tgrav}
is not divergence-free; actually
\begin{equation*}
\nabla\!_a(\Tcal\!\!\spec{grav})_\lambda^a=
\tfrac1G\,\brth^c_\lambda\,(T\Ii b{ca}\,R\iI ba-\tfrac12\,T\Ii b{ad}\,R\iI{bc}{ad})~.
\end{equation*}

Hence we expect that the on-shell divergence
of \hbox{$\Tcal\!\!\spec{gauge}+\Tcal\!\!\spec{matter}$}
depends on the torsion linearly.
Indeed this can be checked, by not-so-short computations.
The vanishing of $\nabla\!_a(\Tcal\!\!\spec{tot})^a_\lambda$ expressed
in terms of the torsion can be regarded as an `integrability condition'
for the gravitational equation.
In the K-G case we obtain
\begin{align*}
\nabla\!_a(\Tcal\!\!\spec{tot})^a_\lambda&=
\tfrac1G\,\brth^c_\lambda\,R\iI{\{a}b\,T\Ii a{c\}b}
+\brth^c_\lambda\,F\!\iIi{ca}ij\,T\Ii{\{e}{be}\,F\Ii{a\}b\,j}i \\
&+\tfrac12\,g^{ae}\,\brth^{\{c}_\lambda\,T\Ii{b\}}{ca}\,
(\nabla\!_e\bar\phi_{\alpha i}\,\nabla\!_b\phi^{\alpha i}
+\nabla\!_b\bar\phi_{\alpha i}\,\nabla\!_e\phi^{\alpha i})\,.
\end{align*}

In the Dirac case we obtain
\begin{align*}
\nabla\!_a(\Tcal\!\!\spec{tot})^a_\lambda&=
\tfrac1G\,\brth^c_\lambda\,
(T\Ii a{cb}\,R\iI ab-\tfrac12\,T\Ii a{bd}\,R\iI{ac}{bd})
\\&\qquad
+\brth^c_\lambda\,F\!\iIi{ca}ij\,
(T\Ii a{eb}\,F\Ii{eb\,j}i -F\Ii{ab\,j}i\,\tau_b)
\\&\qquad
+\tfrac\iO2\,\brth^c_\lambda\,\bigl[\tau_\lambda \gamma\Ii{\lambda\alpha}\beta\,
(\bar\psi_{\alpha i}\,\nabla\!_c\psi^{\beta i}
-\nabla\!_c\bar\psi_{\alpha i}\,\psi^{\beta i})
\\&\qquad\qquad\qquad~
+\bar\psi_{\alpha i}\,
(\gamma^b\,R_{ab}+R_{ab}\,\gamma^b)\Ii\alpha\beta\,\psi^{\beta i}\bigr]\,.
\end{align*}
The last term in the above equation can be further elaborated.
By Clifford algebra we get
\begin{equation*}
\gamma^b\,R_{ab}+R_{ab}\,\gamma^b=
-\tfrac13\,R_{a[bcd]}\,\gamma^b\,\gamma^c\,\gamma^d~,
\end{equation*}
and $R_{a[bcd]}$\,, vanishing in the torsion-free situation,
can be expressed in terms of the exterior covariant differential $\dO[\Gamma]T$,
which is essentially the right-hand side of the first Bianchi equation
with torsion.

\subsubsection{Conclusions}

Offered results support the view that
the tetrad-affine representation of gravity is natural and convenient
under various respects.
In a gauge field theory coupled with gravity there is essentially one
energy-tensor for each sector.
The total energy-tensor is divergence-free,
while the single contributions are not|on account of the torsion.
The torsion itself is unavoidable in this setting,
but it should be regarded as a `byproduct' rather than a fundamental,
independent field.

\end{multicols}

\end{document}